\documentstyle[12pt]{article}
\newcommand{\nn}{\nonumber}
\newcommand{\bd}{\begin{document}}
\newcommand{\ed}{\end{document}}
\newcommand{\bc}{\begin{center}}
\newcommand{\ec}{\end{center}}
\newcommand{\be}{\begin{eqnarray}}
\newcommand{\ee}{\end{eqnarray}}
\newcommand{\ba}{\begin{array}}
\newcommand{\ea}{\end{array}}
\newcommand{\eqn}{\global\def\theequation}
\newcommand{\sw}{sin^2 \theta_W}
\newcommand{\fbd}{f_B}
\renewcommand{\thefootnote}{\alph{footnote}}
\newcommand{\se}{\section}
\newcommand{\sse}{\subsection}
\newcommand{\bi}{\bibitem}
\def\figcap{\section*{Figure Captions\markboth
     {FIGURECAPTIONS}{FIGURECAPTIONS}}\list
     {Figure \arabic{enumi}:\hfill}{\settowidth\labelwidth{Figure 999:}
     \leftmargin\labelwidth
     \advance\leftmargin\labelsep\usecounter{enumi}}}
\let\endfigcap\endlist \relax
\def\reflist{\section*{References\markboth
     {REFLIST}{REFLIST}}\list
     {[\arabic{enumi}]\hfill}{\settowidth\labelwidth{[999]}
     \leftmargin\labelwidth
     \advance\leftmargin\labelsep\usecounter{enumi}}}
\let\endreflist\endlist \relax

\begin{document}
\tolerance=10000
\baselineskip=7mm
\begin{titlepage}

 \vskip 0.5in
 \null
\begin{center}
 \vspace{.15in}
{\LARGE {\bf Study of Radiative Leptonic $D$ Meson Decays }
}\\
\vspace{1.0cm}
  \par
 \vskip 2.1em
 {\large
  \begin{tabular}[t]{c}
{\bf C.~Q.~Geng$^a$, C.~C.~Lih$^{a}$ and Wei-Min Zhang$^{b}$}
\\
\\
{\sl ${}^a$Department of Physics, National Tsing Hua University}
\\  {\sl  $\ $ Hsinchu, Taiwan, Republic of China }
\\
\\
and
\\
\\
       {\sl ${}^b$Department of Physics, National Cheng Kung University}
\\   {\sl  $\ $Tainan, Taiwan,  Republic of China }\\
   \end{tabular}}
 \par \vskip 5.3em
 {\Large\bf Abstract}
\end{center}

We study the radiative leptonic $D$ meson decays of
$D^{+}_{(s)}\to \l^{+}\nu_{\l}\gamma$ ($\l=e,\mu,\tau$), $D^{0}\to
\nu \bar{\nu}\gamma$ and $D^{0}\to \l^{+}\l^{-}\gamma$
($l=e,\mu$) within the light front quark model. In the standard
model, we find that the decay branching ratios of $D^{+}_{(s)}\to
e^{+}\nu_{e}\gamma$, $D^{+}_{(s)}\to \mu^{+}\nu_{\mu}\gamma$ and
$D^{+}_{(s)}\to \tau^{+}\nu_{\tau}\gamma$ are
 $6.9\times 10^{-6}$ ($7.7\times 10^{-5}$),
$2.5\times 10^{-5}$ ($2.6\times 10^{-4}$),
and $6.0\times 10^{-6}$ ($3.2\times 10^{-4}$),
and that of $D^{0}\to \l^{+}\l^{-}\gamma$ ($\l=e,\mu$) and
$D^{0}\to \nu \bar{\nu}\gamma$ are
$6.3\times 10^{-11}$
and $2.7\times 10^{-16}$, respectively.

\end{titlepage}

\se{Introduction}
$\ \ \ $

In the standard model,
flavor changing neutral current (FCNC) decays of $D$ mesons
are predicted to be very rare. Such processes are forbidden
at the tree level but can occur at higher loop level.
On the other hand, these rare decays make a good chance
to probe for new physics \cite{bigi}.
Recently, the new experimental limits with the range
$10^{-4}$ to $10^{-5}$ for FCNC decays of the charm
mesons were obtained at Fermilab E791, E771 and CLEO II
\cite{E791,E771,cleo}.

As we known, the two-body leptonic decays of $D$ mesons of $D\to
\l_i\bar{\l}_j\ (l_{i,j}=e,\mu,\nu)$ are helicity suppressed.
Explicitly, in the standard model, the decay branching ratios of
$D^{+}_{(s)} \to e^{+}\nu_{e}\,,\ \mu^{+}\nu_{\mu}\,,\
\tau^{+}\nu_{\tau}$ and $D^0\to \l^+\l^-$ are $1.0 \times 10^{-8}
\ (9.5 \times 10^{-8})$, $4.7\times 10^{-4}\ (9.5\times 10^{-3})$,
$1.2 \times 10^{-3} \ (4.2 \times 10^{-2})$ and $O(10^{-19})$,
respectively, while that of $D^0 \to \nu\bar{\nu}$ is zero. It is
clear that the rates for the modes of $D\to \l_i\bar{\l}_i\
(l_{i}=e,\mu,\nu)$ are too small to be measured due to the
helicity suppression as well as Glashow-Iliopoulos-Maiani (GIM)
mechanism effect \cite{gim}.

To avoid the helicity suppression, we will consider three body decays
such as $D^+ \to \l^{+}\nu_{l}\gamma$ and
$D^0 \to \l^{+}\l^{-}\gamma$ with
a photon radiated from charged particles.
The amplitudes of the decays
can be divided into the ``internal-bremsstrahlung'' (IB) parts and the 
``structure-dependent'' (SD) parts. 
Since the photon is emission from
the external charged leptons, IB parts are still helicity suppressed for the light charged
lepton modes, while in SD 
one of the photon is emitted from intermediate states
that they are free of the helicity suppression. Therefore, the decay rates of
$D^+ \to \l^{+}\nu_{l}\gamma$ and
$D^0\to l^{+}l^{-}\gamma$ ( $l=e, \mu$ ) might have an enhancement
with respect to the purely leptonic modes of $D^+ \to \l^{+}\nu_{l}$ and
$D^0 \to \l^{+}\l^-$, respectively.

It has been pointed out that the heavy flavor decays can be
analyzed with the help of heavy quark effective theory. For charmed
decays, this is questioned since the charm quark mass is not
much larger than the QCD scale. Therefore, we will use the light front
quark model \cite{r10,wmz,wmz2,wmz3,lf1,lf2} to evaluate the hadronic matrix
elements in $D\to \gamma$ transitions. These decay modes have been
studied in various models \cite{bur,rad1,yan}. It is known that as
the recoil momentum increases, we have to start considering
relativistic effects seriously. The light front quark model
\cite{wmz} is the widely accepted relativistic quark model in
which a consistent and relativistic treatment of quark spins and
the center-of-mass motion can be carried out. In this work, we
calculate form factors in $D\to\gamma$ directly at time-like
momentum transfers by using the relativistic light-front hadronic
wave function. The parameters in the wave function are determined
from Refs. \cite{wmz,wmz2,wmz3}. Within the framework of light-front quark
model for decay processes, one may calculate hadronic matrix elements 
in the frame where the transfer
momentum is purely longitudinal, $i.e$, $p_{\bot}=0$ and
$p^2=p^+p^-$, which covers the entire range of the momentum transfers. 
We will give their 
the momentum dependence in whole kinematics
region of $0\leq p^{2}\leq p_{\max }^{2}$.

The paper is organized as follows. In Sec.~2,
we study the form factors of $D \to \gamma$ transitions
within the light front framework.
We present the decay amplitudes and calculate the
decay branching ratios of $D^+_{(s)} \to \l^{+} \nu_{l}\gamma$,
$D^0 \to \nu \bar{\nu} \gamma$ and $D^0 \to \l^+ \l^- \gamma$
in Sec.~3, 4 and 5, respectively.
In Sec.~6, the conclusion is given.

\section{Form factors of $D\to \gamma$ transitions}

\ \ \

To evaluate the $D_u\to \gamma$ transition , one needs to consider
the following currents:
\be
J^{\mu}&=&\bar{u}\gamma^{\mu}(1-\gamma_{5})c  \nonumber \\
T^{\mu}&=&\bar{u}i\sigma^{\mu\nu}p_{\nu}(1+\gamma_{5})\,c\,,
\ee
where $p$ is the momentum transform, $D_u=\bar{u}c$ and $u$ stands for the light quark
such as a d-quark or u-quark or s-quark
so that $D_d\equiv D^+$, $D_u\equiv D^0$ and $D_s\equiv D^+_s$,
respectively.
Our main task is to calculate the hadronic matrix elements for $D
\to \gamma$ transitions, which can be parametrized as:
\be
\langle\gamma (q_{\gamma})|\bar{u}\gamma_{\mu }\gamma _{5}c|D(p+q_{\gamma})\rangle &=&
-e{\frac{F_{A}}{%
M_{D}}}\left[ (p\cdot q_{\gamma}) \epsilon ^*_\mu
-(\epsilon ^{*}\cdot p)q_{\gamma \mu }\right] ,
\nonumber\\
\langle\gamma (q_{\gamma})|\bar{u}\gamma_{\mu }c|D(p+q_{\gamma})\rangle &=&
ie{\frac{F_{V}}{M_{D}}}\varepsilon_{\mu \alpha \beta \nu }
\epsilon^{*\alpha }p^{\beta }q_\gamma^\nu \, ,
\label{4}
\ee
\be
\langle\gamma(q_{\gamma})|
\bar{u}i\sigma_{\mu\nu}p^{\nu}\gamma_5c|D(p+q_{\gamma})\rangle &=&
ieF_{TA}\epsilon^\nu
(p\cdot q_{\gamma}g_{\mu\nu}-p_\nu q_{\gamma\mu}) ,
\nonumber\\
\langle\gamma(q_{\gamma})|\bar{u}i\sigma_{\mu\nu}p^{\nu}c|D(p+q_{\gamma})\rangle &=&
eF_{TV}\varepsilon_{
\mu\beta\rho\lambda}\epsilon^\beta q_{\gamma}^{\rho} p^\lambda ,
\label{5}
\ee
where the $\epsilon _{\mu }$ is the photon polarization
vector, $q_{\gamma}$ and $p+q_{\gamma}$ are photon and $D$ meson four momenta and
$F_{A}$, $F_{V}$, $F_{TA}$ and $F_{TV}$ are the form factors of
axial-vector, vector, axial-tensor and tensor, respectively.

In the light front approach, the physically accessible kinematics
region is $0\leq p^{2}\leq p_{\max }^{2}=M_{D}^{2}$ due to the time-like
momentum transfers. The $D_u$ meson bound state 
in light-front quark model consists of an
anti-quark $\bar{u}$ and a quark $c$ with the total momentum
$(p+q_{\gamma})$. It can be expressed by:
\begin{eqnarray}
|D(p+q_{\gamma})>&=& \sum_{\lambda _{1}\lambda_{2}}\int [dk_{1}][dk_{2}]
2(2\pi)^{3}\delta ^{3}(p+q_{\gamma}-k_{1}-k_{2})  \nonumber \\
&& ~~~~~~~~ \times \Phi _{D}^{\lambda _{1}\lambda _{2}}(x,k_{\bot})
b_{\bar{u}}^{+}(k_{1},\lambda _{1}) d_{c}^{+}( k_{2},\lambda _{2})
|0>\,,
\end{eqnarray}
where $k_{1(2)}$ is the on-mass shell light front momentum of the internal
quark $\bar{u}(c)$, the light front relative momentum variables $(x,k_{\bot
})$ are defined by
\begin{eqnarray}
k_1^+= x(p+q_{\gamma})^{+}\,,\ k_{1\bot} = x(p+q_{\gamma})_{\bot}+k_{\bot}\,,
\end{eqnarray}
and
\be
\Phi _{D}^{\lambda _{1}\lambda _{2}}(x,k_{\bot })=\left( \frac{%
2k_{1}^{+}k_{2}^{+}}{M_{0}^{2}-\left( m_{u}-m_{b} \right) ^{2}}\right)^{%
\frac{1}{2}}\overline{u}\left( k_{1}, \lambda _{1}\right) \gamma^{5}v\left(
k_{2},\lambda _{2}\right) \phi(x,k_{\bot}) \,,  \label{n6}
\ee
with $\phi(x,k_{\bot})$ being the momentum distribution amplitude. The amplitude
of $\phi(x,k_{\bot})$ can be solved in principles by 
the light-front QCD bound state equation \cite{wmz,wmz2,wmz3}.
However, we use the Gaussian type wave function:
\be
\phi(x,k_{\bot})=N\sqrt{\frac{dk_{z}}{dx}}
\exp \left( -\frac{\vec{k}^{2}} {2\omega_{D}^{2}}\right) \,,
\label{7}
\ee
where
\begin{eqnarray}
& & [dk_1]= {\frac{dk^+dk_{\bot}}{2(2\pi)^3}}\, , \ \ N = 4 \left({\frac{\pi%
}{\omega_{D}^{2}}}\right)^{\frac{3}{4}}\, ,  \nonumber \\
& & k_{z} =\left( x-\frac{1}{2}\right) M_{0}+\frac{m_{c}^{2}-m_{u}^{2}}{%
2M_{0}} \, , \ \ M_0^2={\frac{k^2_{\bot}+m_c^2}{x}}+{\frac{k^2_{\bot}+m_u^2}{%
1-x}} \, ,  \nonumber \\
& & \sum_\lambda u(k,\lambda) \overline{u}(k,\lambda) = {\frac{m + \not{\!
}k }{k^+}} \, , \ \ \sum_\lambda v(k,\lambda) \overline{v}(k,\lambda) = - {%
\frac{m - \not{\! }k }{k^+}} \, .
\end{eqnarray}
For the gauged photon state, one has: \cite{geng2}
\begin{eqnarray}
|\gamma (q_{\gamma})> &=& N^{\prime}\Bigg\{a^{+}(q_{\gamma},\lambda) + \sum_{\lambda_{1}
\lambda_{2}}\int [dk_{1}][dk_{2}]2(2\pi)^{3}\delta^{3} (q_{\gamma}-k_{1}-k_{2})
\nonumber \\
&& ~~~~~~~~~ \times \Phi _{q\bar{q}}^{\lambda_{1}\lambda_{2}\lambda}
(q_{\gamma},k_{1},k_{2}) b_{q}^{+}(k_{1},\lambda _{1})
d_{\bar{q}}^{+} (k_{1},\lambda
_{2}) \Bigg\} | 0 > \, ,
\end{eqnarray}
where
\begin{eqnarray}
\Phi_{q\bar{q}}^{\lambda _{3}\lambda _{4}\lambda }(q_{\gamma},k_{1},k_{2}) &=&\frac{%
e_q}{ED}\chi _{-\lambda _{2}}^{+} \left\{-2\frac{q_{\bot }\cdot \epsilon
_{\bot }} {q^{+}}-\gamma _{\bot}\cdot \epsilon_{\bot } \frac{\gamma _{\bot
}\cdot k_{2_{\bot }}-m_{2}}{k_{2}^{+}} \right.  \nonumber \\
&& ~~~~~~~ \left.-\frac{\gamma _{\bot }\cdot k_{1_{\bot }}-m_{1}} {k_{1}^{+}}%
\gamma_{\bot }\cdot \epsilon _{\bot }\right\} \chi _{\lambda_{1}}\,,
\label{pqq}
\end{eqnarray}
with
\begin{eqnarray}
ED&=&\frac{q_{\gamma \bot }^{2}}{q_{\gamma}^{+}}
-\frac{k_{1_{\bot}}^{2}+m_{1}^{2}} {k_{1}^{+}%
}-\frac{k_{2_{\bot }}^{2}+m_{2}^{2}}{k_{2}^{+}} \,.
\end{eqnarray}
To calculate the matrix elements of the Eqs. (\ref{4}) and (\ref{5}), we choose
a frame with the transverse momentum $p_{\bot}$ = $0$. The 
$p^{2}=p^{+}p^{-} \geq 0$ covers the entire range of the momentum
transfers. By considering the ``good'' component $\mu=+$, the hadronic
matrix elements of Eqs. (\ref{4}) and (\ref{5}) can be rewritten as:
\be
<\gamma (q_{\gamma})| u_{+}^{+}\gamma_{5}c_{+}|D(p+q_{\gamma})>
        &=&-ie\frac{F_{A}}{2M_{D}}\left( \epsilon
        _{\bot }^{*}\cdot q_{\gamma\bot }\right) p^{+}\,,
                \nonumber \\
<\gamma (q_{\gamma})|u_+^+c_+|D(p+q_{\gamma})>&=&e\frac{F_{V}}{2M_{D}}
        \epsilon ^{ij}\epsilon _{i}^{*}q_{\gamma j}p^{+}\,,
                \label{ff}
\ee
and
\begin{eqnarray}
<\gamma (q_{\gamma})|(u_{+}^{+}\gamma^{0}\gamma_5c_{-}-
u_{-}^{+}\gamma^{0}\gamma_5c_{+})|D(p+q_{\gamma})> &=&eF_{TA}%
\left( \epsilon _{\bot }^{*}\cdot q_{\gamma \bot }\right) \,,  \nonumber \\
<\gamma (q_{\gamma})|(u_{+}^{+}\gamma^{0}c_{-}-
u_{-}^{+}\gamma^{0}c_{+})|D(p+q_{\gamma})>&=&-ieF_{TV} \epsilon ^{ij}\epsilon
_{i}^{*}q_{\gamma j}\,,
\label{fft}
\end{eqnarray}
where $u_{+}(c_{+})$ and $u_{-}(c_{-})$ are the light-front up and down components of
the quark fields, and they have the two-component forms:
\be
q_{+}&=&\left(\ba{c}\chi \\ 0\ea\right) \,,
\ee
and
\be
q_{-}&=&\frac{1}{i\partial^{+}}(i\alpha_{\bot}\cdot\partial_{\bot}+\beta m_{q})q_{+}
  \nonumber \\&=&
\left(\ba{c}0 \\ \frac{1}{\partial^{+}}
(\tilde{{\sigma}_{\bot}}\cdot\partial_{\bot}+m_{q})\chi_{q}\ea\right) \,,
\ee
respectively.

In Eq. (15), $\chi_{q}$ is a two-component spinor field and $\sigma$ is the Pauli matrix.
The form factors $F_{V,A}$ and $F_{TV,TA}$ in Eqs. (\ref{ff}) and
(\ref{fft}) are then found to be:
\be
F_{A}(p^{2}) &=&-4M_{D}
        \int \frac{dx\,d^{2}k_{\bot }}{2(2\pi)^{3}}\Phi
        \left( x',k_{\bot }^{2}\right) {1\over 1-x'}
                \nonumber \\
&&~~~~~~~~ \times \left\{ \frac{2}{3}\frac{m_{b}-Bk_{\bot }^{2}
        \Theta}{m_{c}^{2}+
        k_{\bot}^{2}}-\frac{2}{3}\frac{m_{u}-Ak_{\bot }^{2}\Theta }
        {m_{u}^{2}+k_{\bot }^{2}}  \right\}\,, \label{fffa}
\ee
\be
F_{V}(p^{2}) &=&4M_{D}
        \int \frac{dx\,d^{2}k_{\bot }}{2\left( 2\pi \right) ^{3}}\Phi
        \left( x',k_{\bot }^{2}\right) {1\over 1-x'}
                \nonumber \\
&&\left\{ \frac{2}{3}\frac{m_{c}+(1-x')(m_{c}-m_{u}) k_{\bot }^{2}
        \Theta }{m_{c}^{2}+k_{\bot }^{2}}+\frac{2}{3}\frac{m_{u}-
        x'\left( m_{c}-m_{u}\right) k_{\bot }^{2}\Theta }{m_{u}^{2}
        +k_{\bot }^{2}}\right\}\,, \label{fffv}
\ee
\begin{eqnarray}
F_{TA}(p^{2}) &=&\int \frac{dx\,d^{2}k_{\bot }}{2(2\pi)^{3}}%
\Phi \left( x',k_{\bot }^{2}\right) \nonumber
\\
&&~~~~~~~~ \times \left\{ \frac{2}{3}\frac{A_1+A_2\, k_{\bot }^{2} \Theta}{%
m_{c}^{2}+ k_{\bot}^{2}}+\frac{2}{3}\frac{B_1+B_2\, k_{\bot }^{2}\Theta } {%
m_{u}^{2}+k_{\bot }^{2}} \right\}\,,  \label{fffta}
\end{eqnarray}
\begin{eqnarray}
F_{TV}(p^{2}) &=&-\int \frac{dx\,d^{2}k_{\bot }}{2\left( 2\pi
\right) ^{3}}\Phi \left( x',k_{\bot }^{2}\right)
\nonumber \\
&&~~~~~~~~ \times \left\{ \frac{2}{3}\frac{C_1+C_2\, k_{\bot }^{2} \Theta }{
m_{c}^{2}+k_{\bot }^{2}}\ +\frac{2}{3}\frac{D_1+ D_2\, k_{\bot }^{2}\Theta }
{m_{u}^{2} +k_{\bot }^{2}}\right\}\,,
\label{ffftv}
\end{eqnarray}
where
\be
A &=& (1-2x) x'(m_c-m_u) -2xm_u\,, \nn\\
B &=& 2(1-x) x'm_c+(1-2x) (1-x')m_u\,,\nn\\
A_1 &=& \frac{2}{xx'^{2}(1-x')(1-x)}\Bigg\{
(x'+x-2x'x)\left[x'(x-1)-x(2x-1)\right]k_{\bot}^{2}  \nonumber \\
&&+x\left[(x-x')+2x'x(1-x)\right]m_{c}^{2}
+2x^{2}(1-x')^{2}m_{c}m_{u}\Bigg\}\,,  \nonumber \\
A_2 &=& \frac{2(x-x')}{xx'^{2}(1-x')(1-x)}\Bigg\{
(x'+x-2x'x)(1-2x)k^{2}_\bot  \nonumber \\
&&+2x(1-x')m_{u}m_{c}-x(1-2x)(1-x')^{2}m_{u}^{2}
+x'(1+x'x-2xx'^{2})m_{c}^{2}\Bigg\}\,,  \nonumber \\
B_1 &=& \frac{2}{x'x(1-x)(1-x')^{2}}\Bigg\{
(x'+x-2x'x)(1-2x+2x^{2}-x'x)k_{\bot}^{2}  \nn \\
&&+2xx'(1-x')(1-x)m_u\,m_c+(1-x')
(x'+x-4x'x+2x'x^{2})m_{u}^{2} \Bigg\}\,,  \nn \\
B_2 &=& \frac{2(x-x')}{x'x(1-x)(1-x')^{2}}\Bigg\{
-(x'+x-2x'x)(1-2x)k_{\bot}^{2}
\nn \\
&&-\left[(1-2x)x'^{2}(1-x)m_{c}^{2}
+(1-x')(x'+x-3x'x-2x^{2}(1-x'))m_{u}^{2}\right]\Bigg\}\,,  \nn \\
C_1 &=& \frac{2}{x{x'}^2(1-x')(1-x)}\Bigg\{
(x-x'+xx')(x'+x-2x'x)k_{\bot}^{2}  \nn \\
&&-2x^{2}(1-x')^{2}m_c\,m_u
+x'\left[(x-x')-2(1-x')x^{2}\right]m_{c}^{2}\Bigg\}\,, \nonumber \\
C_2 &=& \frac{2(x-x')}{x{x'}^2(1-x')(1-x)}\Bigg\{
(x'+x-2x'x) k_{\bot}^{2}+x'(1-2x+xx')m_{c}^{2}\nn \\
&& -m_{q}^{2}x(1-x')^{2}-2x(1-x')m_u\,m_c\Bigg\}\,, \nonumber \\
D_1 &=& \frac{2}{x(1-x)x'(1-x')^{2}}\Bigg\{
-(1-x)(1-2x+x')(x'+x-2x'x)k_{\bot}^{2}  \nn \\
&&-\left[(1-x')(x'+x-2x^{2}-2x'x+2x^{2}x')m_{u}^{2}
+ 2{x'}^2(1-x)^2 m_{u}m_{c}\right]\Bigg\}\,,  \nonumber \\
D_2 &=& \frac{2(x-x')}{x'(1-x)x(1-x')^{2}}\Bigg\{
(x'+x-2x'x) k_{\bot}^{2} +x'^{2}(1-x)m_{c}^{2} \nn \\
&&+(1-x')(x'-x-x'x)m_{u}^{2}\Bigg\}\,, \nonumber \\
\Phi (x,k_{\bot}^2) &=& N\left( {\frac{2x(1-x) }{M_0^2-(m_u-m_c)^2}}%
\right)^{1/2} \sqrt{{\frac{dk_{z}}{dx}}}\exp \left( -{\frac{\vec{k}^{2}}{%
2\omega_D^2}}\right)\,,  \nonumber \\
\Theta &=& {\frac{1}{\Phi(x,k_{\bot}^2) }} {\frac{d\Phi(x,k_{\bot}^{2})}{%
dk_{\bot}^2}} \, ,  \nonumber \\
x^{\prime}&=&x\left(1-{\frac{p^2}{M_D^2}}\right),\
\vec{k}=(\vec{k}_{\bot}, \vec{k}_{z}) \,.
\label{ffff}
\ee

To illustrate numerical results, we input $m_{u}=0.3,$
$m_{c}=1.4,$ and $\omega =0.5$ in $GeV$. The form factors of
$F_{V,A}$ and $F_{TV,TA}$ are shown in Figures 1 and 2, respectively.

\section{$D^{+}_{(s)}\to \l^{+}\nu_{\l}\gamma$}

\ \ \

\sse{Matrix elements and kinematics}

~~~

In this work, we investigate the decays of $D^{+}_q\to \l^{+}\nu_{\l}\gamma$ $(q=d,s)$.
The effective Hamiltonian for $c\to q\, l\nu_l$
at the quark level in the standard model has been obtained by using similar results
for $b\to q\, l\nu_l$
\be
H_{eff}(c\to q\, l\nu_l)= {G_F\over \sqrt{2}}V_{cq}\bar{q}
\gamma_{\mu}(1-\gamma_5)c\,\bar{\nu}_{\l}\gamma_{\mu}(1-\gamma_5)\l\,.
\label{he1}
\ee
 From Eq. (\ref{he1}), due to the emissions of real photons that
 the decay amplitude of $D^{+}_q\to \l^{+}\nu_{\l}\gamma$ can be written as
\be
M &=&M_{IB}+M_{SD}, \\
M_{IB} &=&i{\frac{G_{F}}{\sqrt{2}}}V_{cq}f_{D}m_l\epsilon _{\mu
}^{*}D^{\mu }, \\
M_{SD} &=&-i{\frac{G_{F}}{\sqrt{2}}}V_{cq}\epsilon _{\alpha }L_{\beta
}H^{\alpha \beta }\,,
\label{He1}
\ee
where $M_{IB}$ and $M_{SD}$ represent the amplitudes from
the IB and SD contributions, respectively,
and
\be
D^{\mu } &=&\bar{u}(p_{\nu })(1+\gamma _{5})(\frac{p^{\mu }}{p\cdot q}-%
\frac{2p_{l}^{\mu }+\not{\! }q\gamma ^{\mu }}{2p_{l}\cdot q}%
)v(p_{l},s_{l}),
\label{25}
\\
L_{\beta } &=&\bar{u}(p_{\nu })\gamma _{\beta }(1-\gamma _{5})v(p_{l},s_{l}),
\label{26}
\\
H^{\alpha \beta } &=&e\frac{F_{A}}{M_{D}}(-g^{\alpha \beta }p\cdot
q+p^{\alpha }q^{\beta })+ie\frac{F_{V}}{M_{D_u}}\epsilon ^{\alpha \beta \mu
\nu }q_{\mu }p_{\nu }\,,
\label{27}
\ee
with $f_{D_{(s)}}$ being the $D_{(s)}$ decay constants,
$\epsilon_{\mu }$ and $s_l$ the four vectors of the photon polarization
and charged lepton, and $p$, $p_{l}$, $p_{\nu}$, and $q$
 the four momenta of $D^{+}_{(s)}$, $l$,
$\nu $ and $\gamma$, respectively.
The physics allowed region of $D^{+}_{(s)}\to \l^{+}\nu_{\l}\gamma$ is
\be
m_{l}^{2}\leq p^{2}\leq M_{D_{(s)}}^{2}\,.
\ee
To describe the kinematics of $D^{+}_{(s)}\to \l^{+}\nu_{\l}\gamma$,
one needs two variables, for
which we define $x_{\gamma}=2E_{\gamma}/M_{D_{(s)}}$ and $y=2E_l/M_{D_{(s)}}$.
We write the physics region for $x_{\gamma}$ and $y$ as:
\be
0 &\leq &x_{\gamma}\leq 1-r_l \,,
\\
1-x_{\gamma}+\frac{r}{1-x_{\gamma}} &\leq &y \leq 1+r_l\,,
\ee
where
\be
r_l=\frac{m_{\l}^2}{M_D^2}= \left\{\ba{cr}
7.1 \times 10^{-8}
&\quad\mbox{for \,\, $\l=e$}\\
3.2 \times 10^{-3}
&\quad\mbox{for \,\, $\l=\mu$}\\
0.9
&\quad\mbox{for \,\, $\l=\tau$} \ea \right.
\ee
for $D^{+}\to \l^{+}\nu_{\l}\gamma$ and
\be
r_l=\frac{m_{\l}^{2}}{M_{D_{s}}^{2}}= \left\{\ba{cr}
6.4 \times 10^{-8}
&\quad\mbox{for \,\, $\l=e$}\\
2.8\times 10^{-3}
&\quad\mbox{for \,\, $\l=\mu$}\\
0.8
&\quad\mbox{for \,\, $\l=\tau$} \ea \right.
\ee
for $D^{+}_s\to \l^{+}\nu_{\l}\gamma$.

\sse{Decay rates}

\ \ \

In the $D^{+}_q\ (q=d$ or $s)$ rest frame, the partial decay rate
is found by \cite{pdg}
\be
d\Gamma =\frac{1}{(2\pi )^{3}}\frac{1}{8M_{D_q}}\mid M\mid ^{2}dE_{\gamma
}dE_{\l}\,.
\label{42}
\ee
Using $x_{\gamma}$ and $y$,
from Eq. (\ref{42}) we obtain the differential decay rate as
\be
\frac{d^{2}\Gamma ^{l}}{dx_{\gamma}dy }
&=&\frac{M_{D}}{256\pi
^{3}}\left| M\right| ^{2}=\frac{M_{D_q}}{256\pi
^{3}}e^{2}G_{F}^{2}V_{cq}^{2}(1-\lambda ) A,
\ee
and
\be
A=A_{IB}(x_{\gamma},y)+A_{SD}(x_{\gamma},y)+A_{IN}(x_{\gamma},y)\,,
\label{43}
\ee
\be
A_{IB}(x_{\gamma},y) &=&
{\frac{4m_{l}^{2}|f_{D_q}|^{2}}{\lambda x_{\gamma}^{2}}}\left[ x_{\gamma}^{2}+2(1-r_l)
\left(1-x_{\gamma}-{\frac{r_l}{\lambda}}\right)\right]\,,
\label{45}
\ee
\be
A_{SD}(x_{\gamma},y) &=&
M_{D_q}^{4}x_{\gamma}^{2}\left[ |F_{V}+F_{A}|^{2}{\frac{\lambda ^{2}}{1-\lambda }}
\left( 1-x_{\gamma}-{\frac{r_{l}}{\lambda }}\right) \right.   \nonumber \\
&&+\left. |F_{V}-F_{A}|^{2}(y-\lambda
)\right]\,,
\label{46}
\ee
\be
A_{IN}(x_{\gamma},y) &=&
-4M_{D_q}m_{l}^{2}\left[ Re[f_{D_q}(F_{V}+F_{A})^{*}]\left( 1-x_{\gamma}-{\frac{
r_l}{\lambda }}\right) \right.   \nonumber \\
&&-\left. Re[f_{D_q}(F_{V}-F_{A})^{*}]{\frac{1-y+\lambda }{\lambda }}
\right]\,,
\label{47}
\ee
where $\lambda =(x_{\gamma}+y-1-r)/x_{\gamma}$.
In Figures 3 and 4, we show the differential branching ratios of
$D^{+}\to \mu^+\nu_{\mu} \gamma$ and $D_{s}^{+}\to \mu^+\nu_{\mu} \gamma$
as  functions of the photon energies, where
we have used $m_d=0.3\ GeV$, $m_s=0.4\ GeV$,
$m_c=1.5\ GeV$, $|V_{cd}|\simeq 0.22$, $|V_{cs}|\simeq 0.974$,
$\omega=0.5\ GeV$, $\tau_{D^{+}_{s}}\simeq 0.467\ ps$,
$\tau_{D^{+}}\simeq 1.05\ ps$ and $f_{D}=f_{D_s}=230\ MeV$ \cite{pdg,fd}.
We get the integrated
branching ratios of $D^{+}$ and $D_{s}^{+}\to l^+\nu_{l}\gamma\
(l=e,\,\mu,\,\tau)$ in Tables 1 and 2.

\begin{table}[h]\caption{Integrated branching ratios for the radiative
leptonic $D^{+}$ decays}
\begin{center}
\begin{tabular}{|c|c|c|c|c|c|}
\hline
Integrated Decay  &  &&&&Ref.\\
Branching Ratios & IB & SD & IN & Sum  &
\cite{yan}  \\ \hline
\hline
$10^{6}B(D^+\to e^{+}\nu_{e} \gamma)$ & $1.0\times 10^{-3}$ & $6.9$ &
$6.2\times 10^{-5}$ & $6.9$ &$82$ \\ \hline
$10^{6}B(D^+\to \mu^+\nu_{\mu} \gamma)$ & $17.2$ & $6.8$ &
$6.2\times 10^{-1}$ & $24.6$ &$-$ \\ \hline
$10^{6}B(D^+\to \tau^+\nu_{\tau} \gamma)$ & & &  && \\
(Cut $\delta=0.01$) & $5.6$ &
$1.3\times 10^{-6}$ & $3.7\times 10^{-1}$ & $6.0$ &$-$ \\ \hline
\end{tabular}\end{center}\end{table}

\begin{table}[h]\caption{Integrated branching ratios for the radiative
leptonic $D_{s}^{+}$ decays}
\begin{center}
\begin{tabular}{|c|c|c|c|c|c|c|}
\hline
Integrated Decay  &  &&&&Ref.&Ref.\\
Branching Ratios & IB & SD & IN & Sum  &
\cite{yan} &
 \cite{rad1}  \\ \hline
\hline
$10^{5}B(D_{s}^{+}\to e^{+}\nu_{e} \gamma)$ & $1.1\times 10^{-3}$ & $7.7$ &
$6.8\times 10^{-5}$ & $7.7$ &$90$ &$17$ \\ \hline
$10^{5}B(D_{s}^{+}\to \mu^+\nu_{\mu} \gamma)$ & $17.8$ & $7.5$ &
$7.1\times 10^{-1}$ & $26.0$ &$-$ &$17$ \\ \hline
$10^{5}B(D_{s}^{+}\to \tau^+\nu_{\tau} \gamma)$ & & &  &  && \\
(Cut $\delta=0.01$) & $30.7$ &
$1.8\times 10^{-3}$ & $1.7$ & $32.4$ &$-$& $-$ \\ \hline
\end{tabular}\end{center}\end{table}

We now compare our results with Refs. \cite{rad1,yan}.
As shown in Tables 1 and 2, our result are much smaller
than that in Ref. \cite{yan} but close to
the values in Ref. \cite{rad1}.
We note that  for the $\mu$ and $\tau$ channels
the IB contributions are dominant since
D-mesons are not very heavy.

\section{ $D^{0}\to \nu \bar{\nu}\gamma$}

\ \ \

\sse{Matrix elements and kinematics}

For the processes of $D^{0}\to \nu_l\bar{\nu}_l\gamma\
(l=e,\mu,\tau)$, at the quark level, they arise from the box and
$Z$-penguin diagrams
that contribute to $c\to u\nu_l\bar{\nu}_l$ with the photon
emitting from the charged particles in the diagrams. However, when the
photon line is attached to the internal charge lines such as the W boson and
quark lines, there is a suppression factor of $m_c^2/M_W^2$ in the
Wilson coefficient in comparing with those in $c\to
u\nu_l\bar{\nu}_l$.
Thus, we need only to consider the diagrams with
the photon from the external quarks.
The effective Hamiltonian for $c\to u\,\nu_l\,\bar{\nu}_l$ is given by
\be
H_{eff}(D^{0}\to \nu_l\bar{\nu}_l\gamma)=
{G_F\over \sqrt{2}}{\alpha\over
2\pi\sin^2\theta_W}\sum_{i=d,s,b}
V_{ui}V_{ci}^{*}D(x_i)\bar{u}
\gamma_{\mu}(1-\gamma_5)\,c\bar{\nu}_l\gamma_{\nu}(1-\gamma_5)\nu_l\,,
\label{He2}
\ee
where
\be
D(x_i)&=& {1\over 4}\left[\,{x^{2}_{i}-x_i+3\over x_i-1}
+{3(x_i-2)\over (1-x_i)^2}x_i\ln
x_i\right]\,,
\label{Dxt}
\ee
and $x_i=m_i^2/M_W^2$ with $m_i\ (i=d,s,b)$ being the current quark masses.
We note that in Eqs. (\ref{He2}) and (\ref{Dxt}),
only the leading contributions have been included and the additional
$\alpha_s$ correction to the result, which are small \cite{lim}.

\sse{Decay amplitude}
 From the effective Hamiltonian for
$c\to u\,\nu_l\,\bar{\nu}_l$ of Eq. (\ref{He2}) and
the form factors defined in Eq. (\ref{ff}),
we can write the amplitude of $D^0 \to \nu_l\bar{\nu}_{l}\gamma$ as
\be
M &=& -ie
{G_F\over \sqrt{2}}{\alpha\over
2\pi\sin^2\theta_W}\sum_{i=d,s,b}
V_{ui}V_{ci}^{*}D(x_i)
\epsilon_{\mu }^{*}H^{\mu \nu }\bar{u}(p_{\bar{\nu}}) \gamma
_{\mu }(1-\gamma_{5})v(p_{\nu})\,,
\ee
with
\be
H_{\mu \nu } &=&{F_A\over M_D}(-p\cdot q_{\gamma} \,g_{\mu\nu}+p_{\mu }q_{\gamma \nu })
+i\epsilon _{\mu \nu \alpha \beta }\frac{F_{V}}{M_D}q^{\alpha}_{\gamma}p^{\beta}\,.
\ee
where $M_D$ is the mass of $D^0$ and
the form factors are given by Eqs. (\ref{fffa}) and (\ref{fffv}).

Similar to the decays discussed in the previous subsection,
we also define $x_{\gamma}=2E_{\gamma}/M_D$ and $y=2E_{\bar{\nu}}/M_D$
in the $D$-meson rest frame in order to re-scale the energies of the
photon and anti-neutrino.
By integrating the variable $y$ in the
phase space, we obtain the differential decay
rate of $D^0\to \nu\bar{\nu}\gamma$ as
\be
\frac{d\Gamma }{dx_{\gamma}} &=& 2\alpha \left({G_F\alpha\over
16\pi^2\sin^2\theta_W}\right)^2
(|F_A|^2+|F_V|^2)\sum_{i=d,s,b}|V_{ui}V_{ci}^{*}|^2
D^2(x_i)x_{\gamma}^{3}(1-x_{\gamma})
M_{D}^{5}\,,
\label{tdr}
\ee
where we have included the three generations of neutrinos.

Using $m_c=1.4\ GeV$, $m_u=300\ MeV$, and $\omega=0.5$, the differential decay
branching ratio $dB(D^0\to \nu\bar{\nu }\gamma)/dx_{\gamma}$ as a function of
$x_{\gamma}=2E_{\gamma}/M_D$ is shown in Figure 5 and the integrated decay
branching ratio is
\be
B(D^{0}\to \nu \bar{\nu}\gamma ) &=&2.7\times 10^{-16} \,.
\ee

It is easy to see that in the standard model the branching ratio
of $D^{0}\to \nu \bar{\nu}\gamma$ is too small to be measured
experimentally. However, a large branching ratio may be induced by
new physics such as the leptoquark model \cite{genglq}.

\section{ $D^{0}\to \l^{+}\l^{-}\gamma$}

\ \ \

\sse{Matrix elements and kinematics}

The contribution to the process of $D^{0} \to \l^{+} \l^{-}\gamma$
($l=e$ or $\mu$) arises from the effective Hamiltonian
that induces the pure leptonic mode of
$D^{0} \to l^{+}l^{-}$. The short distance contribution for $c \to u\,\l^{+} \l^{-}$
comes from the $W$-box, Z-boson and
photon penguin diagrams, and the
effective Hamiltonian is given by
\be
{\cal L}&=&-\frac{G_{F}\alpha}{2\sqrt{2}\pi\sin^{2}{\theta_{W}}}\sum_{i=d,s,b}
V_{ui}V_{ci}^{*}
\Bigg\{F^{1}_{i}\bar{u}\gamma_{\mu}P_{L}c\,\bar{l}\gamma^{\mu}(1-\gamma_{5})l+F^{2}_{i}
\bar{u}\gamma_{\mu}P_{L}c\,\bar{l}\gamma^{\mu}(1+\gamma_{5})l
 \nn\\
 && ~~~~~~~~
-2\frac{F^{3}_{i}\sin^{2}{\theta_{W}}}{p^2}m_{c}\bar{u}\,i\sigma_{\mu\nu}p^{\nu}P_{R}c\,
\bar{l}\gamma^{\mu}l\Bigg\} ,
\label{2}
\ee
where $P_{L,R}$ =$(1\mp \gamma_5)/2$,
and $p$ is the momentum transfer which is
equal to the momentum of the lepton pair. The Wilson coefficients
$F^{1}_{i}, F^{2}_{i}, F^{3}_{i}$ can be found in Refs.
\cite{lim,aj} and one has
\be
F^{1}_{i}&=&C^{1}_{i}+C^{2}_{i}-\sin ^2\theta_W(C^{3}_{i}+2C^{2}_{i})~,\nonumber\\
F^{2}_{i}&=&-\sin ^2\theta_W(C^{3}_{i}+2C^{2}_{i})\,,
\nonumber\\
F^{3}_{i}&=&-Q\Biggl(\biggl[-{1\over 4}{1\over x_i-1}+{3\over 4}{1\over
(x_i-1)^2}+{3\over 2}{1\over (x_i-1)^3}\biggr]x_i-{3\over 2}{x_i^2\ln x_i\over
(x_i-1)^4}\Biggr)\nonumber\\
&+&\biggl[{1\over 2}{1\over x_i-1}+{9\over 4}{1\over (x_i-1)^2} +{3\over
2}{1\over (x_i-1)^3}\biggr]x_i-{3\over 2}{x_i^3\ln x_i\over (x_i-1)^4}\,,
\ee
where $x_i=m_i^2/M_W^2$ with $m_i\ (i=d,s,b)$ being the current quark masses.
The coefficients of $C^{1}_{i}$, $C^{2}_{i}$ and $C^{3}_{i}$
are given by
\be
C^{1}_{i}&=&{3\over 8} \biggl[-{1\over x_i-1}+{x_i\ln x_i \over
(x_i-1)^2}\biggr]-\gamma (\xi,x_i)\,,
\nonumber\\
C^{2}_{i}&=&{x_i\over 4}-{3\over 8}{1\over x_i-1}+{3\over 8}{2x_i^2-x_i\over
(x_i-1)^2}\ln x_i +\gamma (\xi,x_i)\,,
\nonumber\\
C^{3}_{i}&=&Q\Biggl(\biggl[{1\over 12}{1\over x_i-1}+{13\over 12}{1\over
(x_i-1)^2}-{1\over 2}{1\over (x_i-1)^3}\biggr]x_i\nonumber\\
&+&\biggl[{2\over 3}{1\over x_i-1}+\biggl({2\over 3}{1\over (x_i-1)^2}-{5\over
6}{1\over (x_i-1)^3}+{1\over 2}{1\over (x_i-1)^4}\biggr)x_i\biggr]\ln
x_i\Biggr)\nonumber\\
&-&\biggl[{7\over 3}{1\over x_i-1}+{13\over 12}{1\over (x_i-1)^2}-{1\over
2}{1\over (x_i-1)^3}\biggr]x_i\nonumber\\
&-&\biggl[{1\over 6}{1\over x_i-1}-{35\over 12}{1\over (x_i-1)^2}-{5\over
6}{1\over (x_i-1)^3}+{1\over 2}{1\over (x_i-1)^4}\biggr]x_i\ln x_i-2\gamma
(\xi,x_i)\,,
\ee
and
\be
\gamma(\xi,x_{i})&=&{1\over \xi x_{i}-1}\Biggl({3\over 4}{1\over x_i-1}+{1\over 8}
{1\over \xi x_{i}-1}\Biggr)x_i\ln x_i-{1\over 8}{1\over \xi}{1\over \xi x_i-1}
\nonumber\\
&\times &\Biggl[\biggl({5\xi +1\over \xi -1}-{1\over \xi x_i-1}\biggr)\ln \xi
+1\Biggr]\,,
\ee
where $\xi$ is a gauge dependent parameter.
The decay amplitude for $D^0 \to l^{+}l^{-}\gamma $ can be written as
\be
{\cal M}(D^0\to l^{+}l^{-}\gamma)={\cal M}_{IB}+{\cal M}_{SD}
\label{2.1}
\ee
where ${\cal M}_{IB}$ and ${\cal M}_{SD}$ represent the $IB$ and $SD$ contributions,
respectively.
For the $IB$ part, the amplitude is clearly proportional to the
lepton mass $m_l$ and it can be written as:
\be
{\cal
M}_{IB}&=&-ie\frac{G_{F}\alpha}{2\sqrt{2}\pi\sin^{2}{\theta_{W}}}
\sum_{i=d,s,b}V_{ui}V_{ci}^{*}f_{D}m_{\l}(F_{i}^{2}-F_{i}^{1})
\left[\bar{\l}(\frac{\not{\! }\epsilon\not{\!
}P_{D}}{2p_{1}\cdot q_{\gamma}} -\frac{\not{\! }P_{D}\not{\!
}\epsilon}{2p_{2}\cdot q_{\gamma}})\gamma_{5}\l\right] , \label{3.1} \ee
where $P_D$, $p_1$, $p_2$ and $q_{\gamma}$, are the momenta of
$D$, $\l^+$, $\l^-$ and photon respectively. The
amplitude for the $SD$ part, where a photon is from the initial quark line,
is given by
\be
{\cal M}_{SD}&=&-\frac{G_{F}\alpha}{2\sqrt{2}\pi\sin^{2}{\theta_{W}}}
\sum_{i=d,s,b}V_{ui}V_{ci}^{*}
\Bigg\{F_{i}^{1}
\langle\gamma(q_{\gamma})|\bar{u}\gamma_{\mu}P_{L}c|D(p+q_{\gamma})\rangle
\bar{l}\gamma^{\mu}(1-\gamma_{5})l
 \nn\\
 && ~~~~~~~~
+F_{i}^{2}
\langle\gamma(q_{\gamma})|\bar{u}\gamma_{\mu}P_{L}c|D(p+q_{\gamma})
\rangle\bar{l}\gamma^{\mu}(1+\gamma_{5})l
 \nn\\
 && ~~~~~~~~
-2\frac{F_{i}^{3}m_c\,\sin^{2}{\theta_{W}}}{p^2}
\langle\gamma(q_{\gamma})|\bar{u}\,i\sigma_{\mu\nu}p^{\nu}P_{R}c|D(p+q_{\gamma})\rangle
\bar{l}\gamma^{\mu}l\Bigg\} ,
\label{3.2}
\ee
 which shows that to find the decay rate,
one has to evaluate the hadronic matrix elements in Eq. (\ref{3.2}). 

 Using Eqs. (2) and (3), we rewrite Eqs. (\ref{3.2}) as:
\be
{\cal M}_{SD} &=& \frac{G_{F}\alpha}{2\sqrt{2}\pi\sin^{2}{\theta_{W}}}
\sum_{i=d,s,b}V_{ui}V_{ci}^{*}
\Bigg{\{} \epsilon_{\mu \alpha \beta \sigma} \epsilon^{*\alpha} p^\beta
q_{\gamma}^{\sigma} \left[ A \, \bar \l \gamma^\mu \l + C \, \bar \l \gamma^\mu
\gamma_5 \l \right] ~
\nn \\
&& +~  i \left[ \epsilon_\mu^* (p\cdot q_{\gamma}) - (\epsilon^* \cdot p )
q_{\gamma\,\mu} \right]
\left[ B \, \bar \l \gamma^\mu \l + D  \, \bar \l \gamma^\mu \gamma_5
\l \right] \Bigg{\}}~,
\label{5.1}
\ee
where the factors of $A$-$D$ are
\be
A &=& \frac{(F_{i}^{1}+F_{i}^{2})}{M_{D}} F_{VA}(p^2)
-2\,F_{i}^{3}\frac{m_c}{p^{2}}F_{TA}(p^2)~, \nn \\
B &=& \frac{(F_{i}^{1}+F_{i}^{2})}{M_{D}} F_{VV}(p^2)
-2\,F_{i}^{3}\frac{m_c}{p^{2}}F_{TV}(p^2)~, \nn \\
C &=& \frac{(F_{i}^{2}-F_{i}^{1})}{M_{D}} F_{VA}(p^2), \nn \\
D &=& \frac{(F_{i}^{2}-F_{i}^{1})}{M_{D}} F_{VV}(p^2),
\ee
respectively.

\sse{Decay rates}

The partial decay width for $D^{0} \to l^+ l^- \gamma$ in the $D^0$ rest
frame is found to be:
\be
d\Gamma={1\over 2M_D}|{\cal M}|^2 (2\pi)^4\delta(P-p_{1}-p_{2}-q_{\gamma})
{d\vec{q}\over (2\pi)^3 2E_{\gamma}} {d\vec{p}_{1}\over (2\pi)^3
2E_{1}} {d\vec{p}_{2}\over (2\pi)^3 2E_{2}},
\ee
where the square of the matrix element is given by
\be
|{\cal M}|^2=|{\cal M}_{IB}|^2+|{\cal M}_{SD}|^2+2Re({\cal
M}_{IB}{\cal M}_{SD}^{*})\,.
\ee
To describe the kinematics of
$D^{0} \to l^+ l^- \gamma$, we define two variables of
$x_{\gamma}=2P_{D}\cdot q_{\gamma}/M_D$ and $y=2P_{D}\cdot
p_{1}/M_D$. We write the transfer momentum $p^2$ in term of
$x_{\gamma}$ as:
\be
p^2=M_{D}^{2}(1-x_{\gamma}).
\ee
The double differential decay rate is then found to be:
\be
\frac{d^{2}\Gamma ^{l}}{dx_{\gamma}d\lambda } &=&\frac{M_{D}}{256\pi
^{3}}\left| M\right| ^{2}=C\rho (x_{\gamma},\lambda) ,
\ee
where
\be
C=\alpha\sum_{i=d,s,b}|\frac{\alpha V_{ui}V_{ci}^{*}}{8\pi^{2}}|^{2}G_{F}^{2}M_{D}^{5}
\ee
and
\be
\rho (x_{\gamma},\lambda) &=&\rho_{IB}(x_{\gamma},\lambda)
+\rho_{SD}(x_{\gamma},\lambda)+\rho_{IN}(x_{\gamma},\lambda),
\ee
with
\be
\rho_{IB} &=&4|f_{D}(F_{i}^{2}-F_{i}^{1})|^{2}\frac{r_l}{M_{D}^{2}x_{\gamma}^{2}}\Bigg\{
\frac{x_{\gamma}^{2}-2x_{\gamma}+2-4r_{l}}{\lambda(1-\lambda)}-2r_{l}(
\frac{1}{\lambda^{2}}+\frac{1}{(1-\lambda)^{2}})\Bigg\},  \nonumber \\
\rho_{SD} &=&\frac{M_{D}^{2}}{8}x_{\gamma}^{2}\Bigg\{(|A|^{2}+|B|^{2})
\left[(1-x_{\gamma}+2r_{l})
-2(1-x_{\gamma})(\lambda-\lambda ^{2})\right]  \nn \\
&&+(|C|^{2}+|D|^{2})\left[(1-x_{\gamma}-2r_{l})
-2(1-x_{\gamma})(\lambda-\lambda ^{2})\right]  \nn \\
&&+2Re(B^*C+A^*D)
(1-x_{\gamma})(2\lambda-1)\Bigg\} ,  \nonumber \\
\rho_{IN} &=&f_{D}(F_{i}^{2}-F_{i}^{1})r_{l}\Bigg\{Re(A)
\frac{x_{\gamma}}{\lambda(1-\lambda)}+Re(D)
\frac{x_{\gamma}(1-2\lambda)}{\lambda(1-\lambda)}\Bigg\}\, .
\ee
Here $\lambda =(x_{\gamma}+y-1)/x_{\gamma}$ and $r_l=m_{l}^2/M_{D}^2$ and
the physical regions for $x_{\gamma}$ and $\lambda$ are given by:
\be
0 &\leq &x_{\gamma}\leq 1-4r_{l} \,,  \nonumber \\
\frac{1}{2}-\frac{1}{2}\sqrt{1-\frac{4r_{l}}{1-x_{\gamma}}} &\leq &\lambda
\leq \frac{1}{2}+\frac{1}{2}\sqrt{1-\frac{4r_{l}}{1-x_{\gamma}}}\,.
\ee

In Figure 6 we present the differential decay rate of $D^0 \to \mu^+
\mu^- \gamma$ as functions of $x_{\gamma}$.
 From the figures we see that the contributions from soft photons, $i.e$ the IB ones,
corresponding to small $x_{\gamma}$ region, are infrared
divergence. To obtain the decay width of $D^0 \to \l^+\l^-
\gamma$, a cut on the photon energy is needed. The integrated
branching ratios are summarized in Table 3. Here, we have used the
cut value of $\delta =0.01$ and $m_c=1.5 GeV$, $m_s=0.4\ GeV$ and
$m_d=0.3\ GeV$ in the calculations of the form factors. We have
also used $\vert V_{cd}V_{ud}^*\vert =\vert V_{cs}V_{us}^*\vert=
0.2145$, $\vert V_{cb} V_{ub}^*\vert = 1.6 \times 10^{-4}$,
$\tau(D^0) = 0.415 \times 10^{-12}~s$ \cite{pdg} to evaluate the
numerical results.

\begin{table}[h]\caption{Integrated branching ratios for the radiative
leptonic $D^0$ decays}
\begin{center}
\begin{tabular}{|c|c|c|c|c|}
\hline
Integrated Decay  &  &&&\\
Branching Ratios & IB & SD & IN & Sum   \\ \hline
\hline
$10^{11}B(D^0\to e^{+}e^{-}\gamma)$ & $1.3 \times 10^{-9}$ & $6.3$ &
$4.6\times 10^{-4}$ & $6.3$ \\ \hline
$10^{11}B(D^0\to \mu^+\mu^-\gamma)$ & $1.9 \times 10^{-5}$ & $6.3$ &
$2.2\times 10^{-3}$ & $6.3$ \\ \hline
\end{tabular}\end{center}\end{table}

We see that the decay branching ratios of the radiative leptonic
decays of $D^0\to \l^{+}\l^{-}\gamma$ are about eight orders of
magnitude larger than that of the purely leptonic modes of $D^0\to
\l^{+}\l^{-}$ for $l=e$ and $\mu$, respectively.

\section{Conclusions}

\ \ \

We have calculated the form factors of the $D \to \gamma$
within the light-front model and used these
form factors to evaluate the branching ratios of the leptonic
$D$-meson radiative decays.
We have found that,
in the standard model, the decay branching ratios
of
$D^{+}_{(s)}\to e^{+}\nu_{e}\gamma$, $D^{+}_{(s)}\to \mu^{+}\nu_{\mu}\gamma$
and $D^{+}_{(s)}\to \tau^{+}\nu_{\tau}\gamma$
are
 $6.9\times 10^{-6}$ ($7.7\times 10^{-5}$),
$2.5\times 10^{-5}$ ($2.6\times 10^{-4}$), and $6.0\times 10^{-6}$
($3.2\times 10^{-4}$), and that of $D^{0}\to \l^{+}\l^{-}\gamma$
($\l=e,\mu$) and $D^{0}\to \nu \bar{\nu}\gamma$ are $6.3\times
10^{-11}$ and $2.7\times 10^{-16}$, respectively. Comparing with
the purely leptonic decays of $D^0 \to l^+ l^-$, we have found
that the rates of the corresponding radiative modes are  several
orders of magnitude larger. We conclude that some of the radiative
leptonic decays could be measured in the BTeV/C0 experiment at
Fermilab \cite{tev}. During Tevatron Run II, which could
reconstruct about $10^{9}$ charm mesons decays and increase the
upper limit of about three orders.

\vspace{1.5cm}

\noindent
{\bf Acknowledgments}

This work was supported in part by the National Science Council of the
Republic of China under contract numbers
 NSC-89-2112-M-007-013 and
NSC-89-2112-M-006-026.

\newpage

\begin{figcap}

\item
The values of the form factors $F_V$ (solid curve) and $F_A$ (dashed
curve) as functions of the momentum transfer $p^2$
for $D^+_s\to \gamma$.
\item
The values of the form factors $F_{TV}$ (solid curve) and $F_{TA}$ (dashed
curve) as functions of the momentum transfer $p^2$
for $D^+_s\to \gamma$.
\item
The differential decay
branching ratio $dB(D^{+}\to \mu^+\nu_{\mu} \gamma)/dx_{\gamma}$ as a
function of
$x_{\gamma}=2E_{\gamma}/M_{D}$.
\item
The differential decay
branching ratio $dB(D_{s}^{+}\to \mu^+\nu_{\mu} \gamma)/dx_{\gamma}$ as a
function of
$x_{\gamma}=2E_{\gamma}/M_{D_{s}}$.
\item
The differential decay
branching ratio $dB(D^{0}\to \nu \bar{\nu}\gamma)/dx_{\gamma}$ as a
function of
$x_{\gamma}=2E_{\gamma}/M_{D}$.
\item
The differential decay
branching ratio  $dB(D^0 \to \mu^+\mu^-\gamma)/dx_{\gamma}$ as a
function of
$x_{\gamma}=2E_{\gamma}/M_{D}$.

\end{figcap}

\newpage
\begin{figure}[h]
\includegraphics{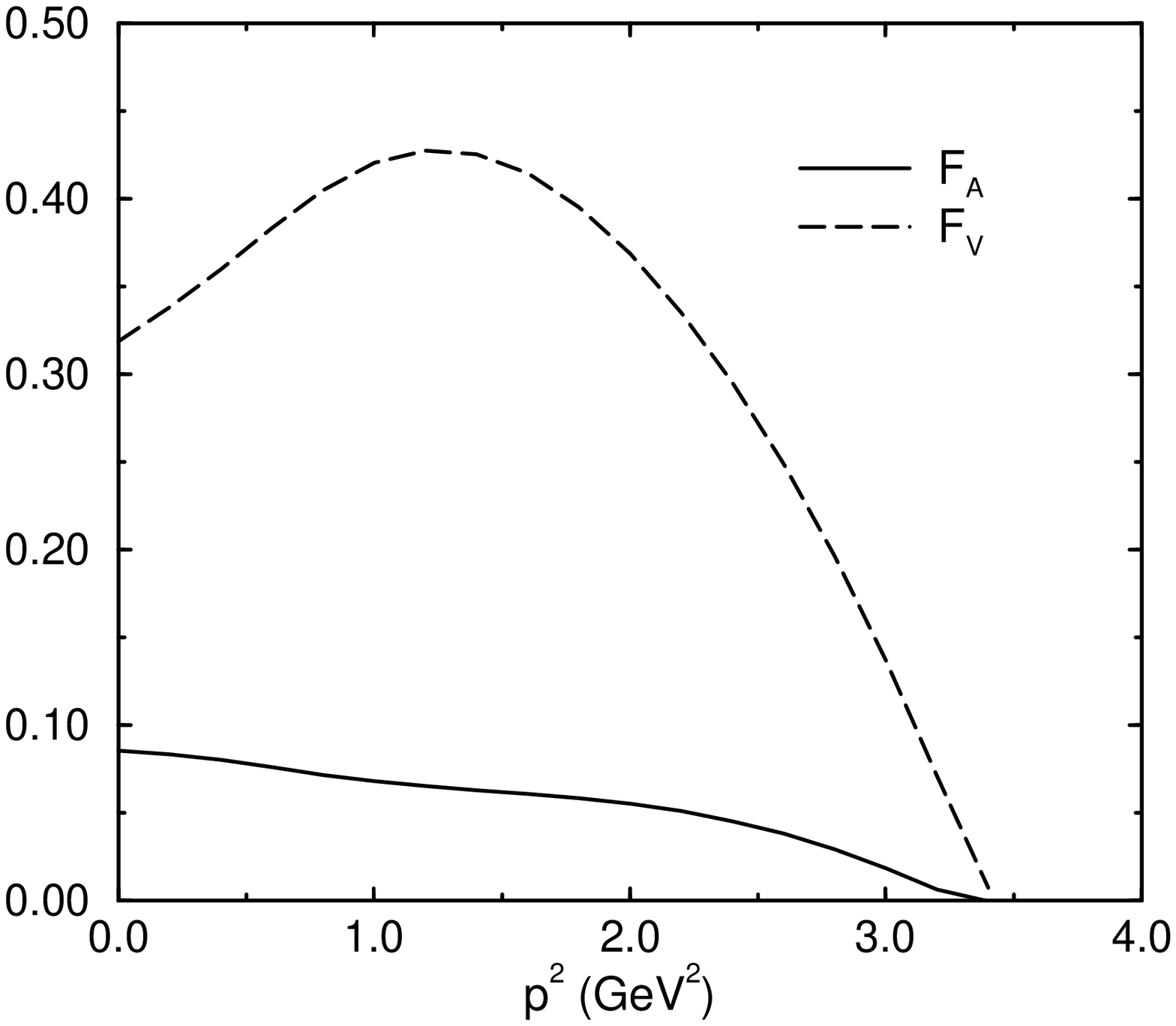}
\vskip 13cm
\caption{
The values of the form factors $F_{A}$ (solid curve) and $F_{V}$ (dashed
curve) as functions of the momentum transfer $p^2$
for $D^{+}_{s}\to \gamma$.
}
\end{figure}

\newpage
\begin{figure}[h]
\includegraphics{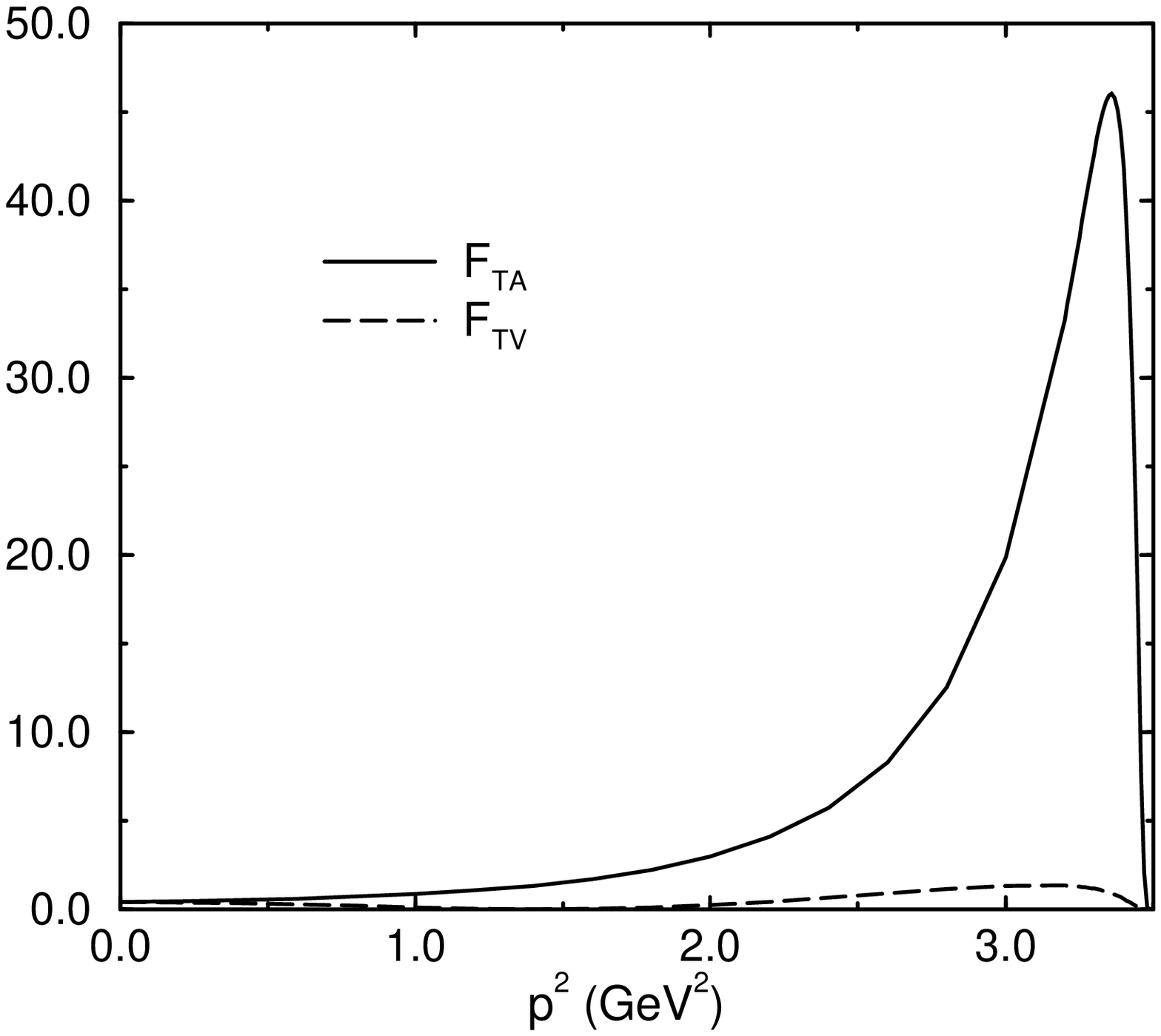}
\vskip 13cm
\caption{
The values of the form factors $F_{TA}$ (solid curve) and $F_{TV}$ (dashed
curve) as functions of the momentum transfer $p^2$
for $D^{+}_{s}\to \gamma$.
}
\end{figure}

\newpage
\begin{figure}[h]
\includegraphics{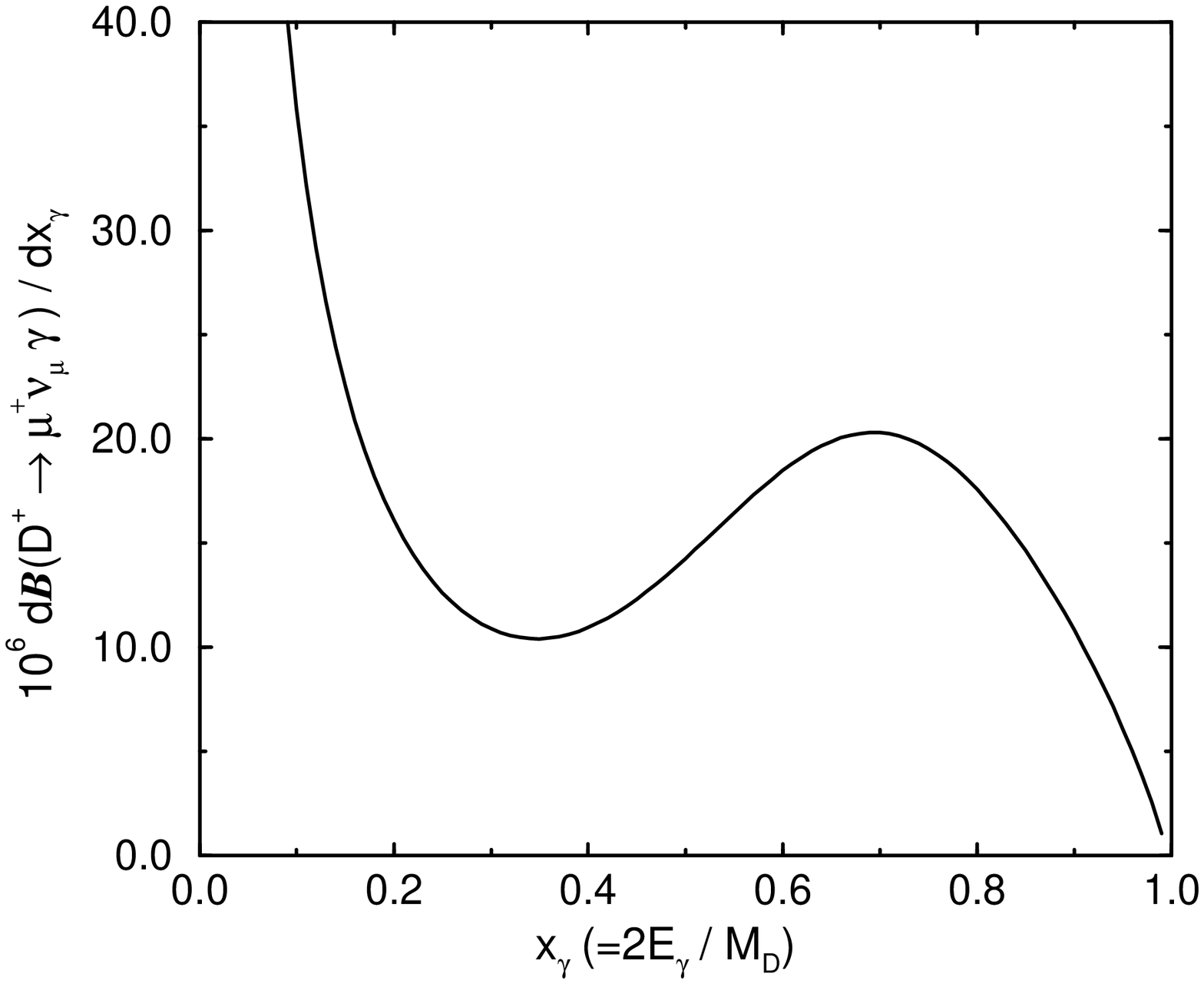}
\vskip 13cm
\caption{The differential decay
branching ratio $dB(D^{+}\to \mu^+\nu_{\mu} \gamma)/dx_{\gamma}$ as a
function of $x_{\gamma}=2E_{\gamma}/M_{D}$.}
\end{figure}

\newpage
\begin{figure}[h]
\includegraphics{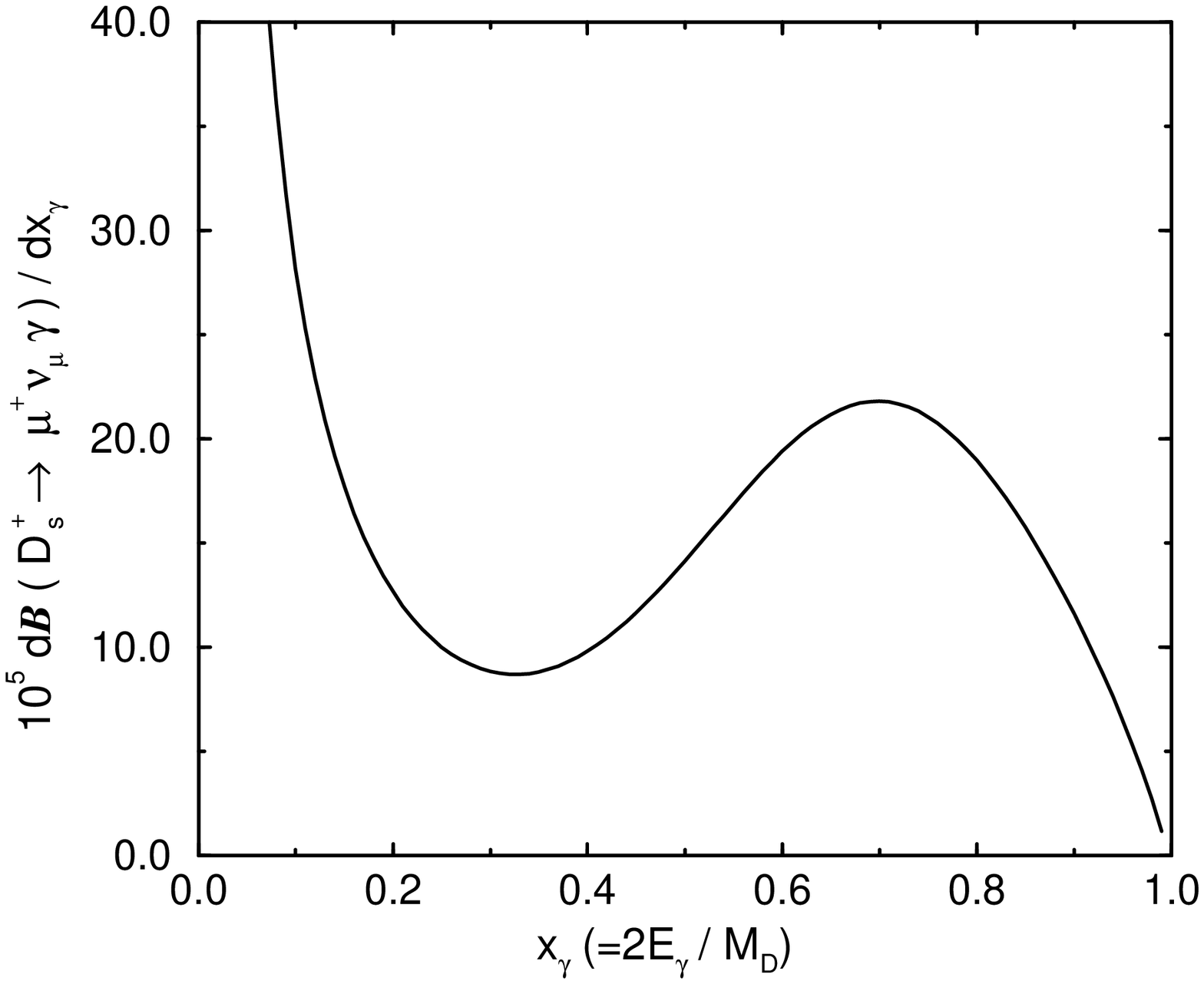}
\vskip 13cm
\caption{The differential decay
branching ratio $dB(D_{s}^{+}\to \mu^+\nu_{\mu} \gamma)/dx_{\gamma}$ as a
function of $x_{\gamma}=2E_{\gamma}/M_{D}$.}
\end{figure}

\newpage
\begin{figure}[h]
\includegraphics{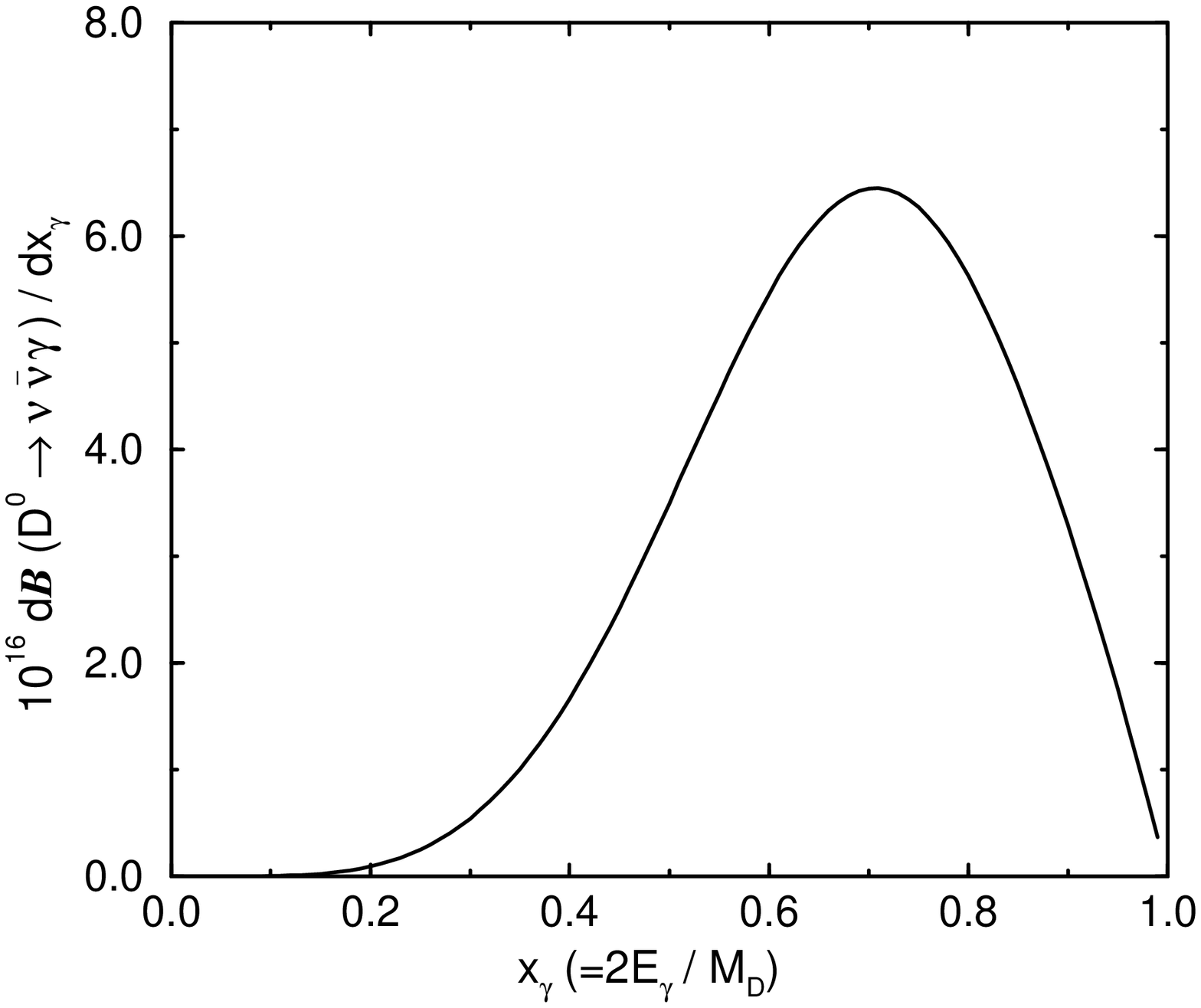}
\vskip 13cm
\caption{The differential decay
branching ratio $dB(D^{0}\to \nu \bar{\nu}\gamma)/dx_{\gamma}$ as a
function of $x_{\gamma}=2E_{\gamma}/M_{D}$.}
\end{figure}

\newpage
\begin{figure}[h]
\includegraphics{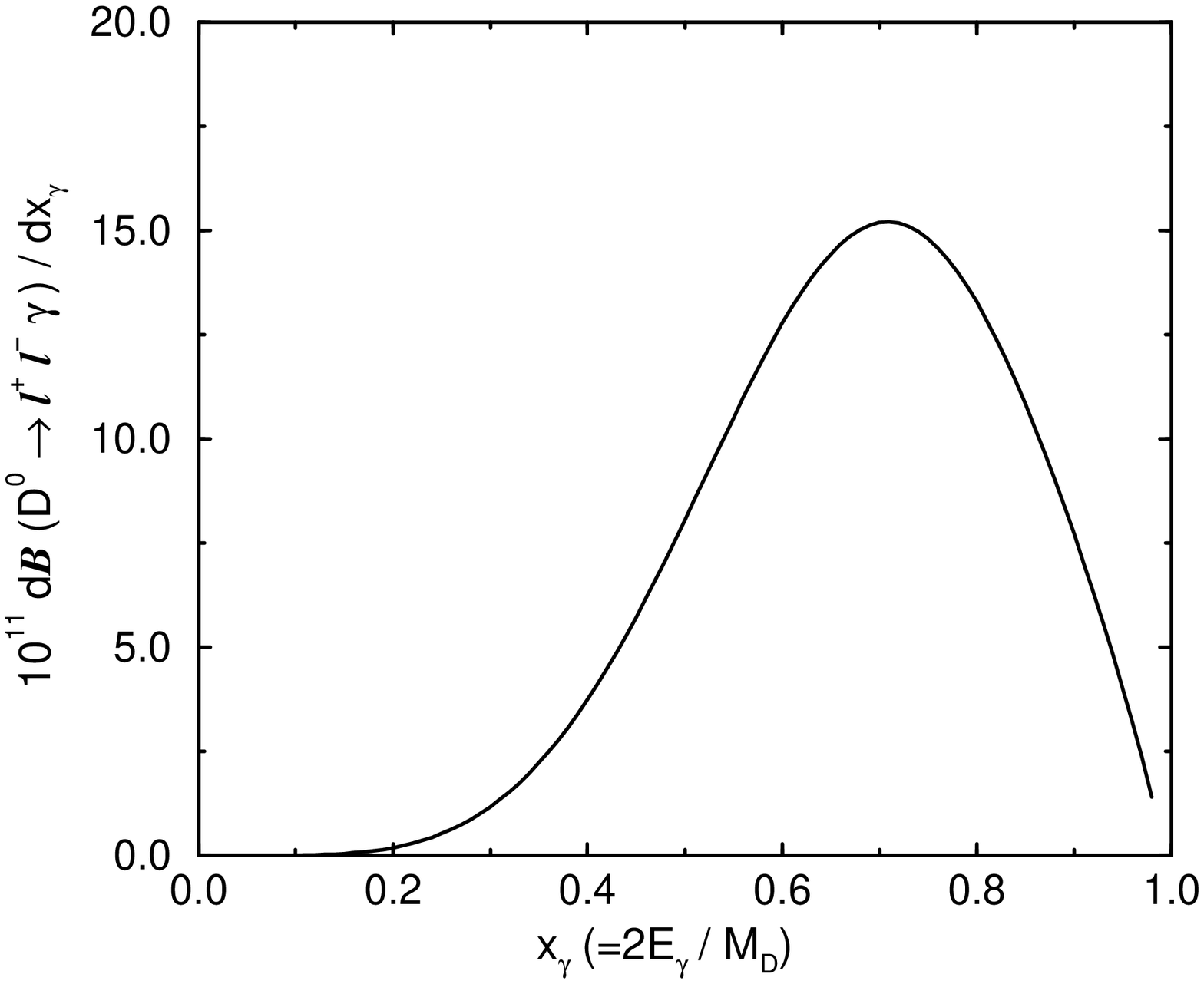}
\vskip 13cm
\caption{The differential decay
branching ratio $dB(D^0 \to \mu^+\mu^-\gamma)/dx_{\gamma}$ as a
function of $x_{\gamma}=2E_{\gamma}/M_{D}$.}
\end{figure}

\end{document}